# Active laser frequency stabilization and resolution enhancement of interferometers for the measurement of gravitational waves in space

Markus Herz[*]

*EADS Astrium GmbH, 88039 Friedrichshafen, Germany.*

---

[*] Email address: markus.herz@astrium.eads.net

# Abstract

Laser frequency stabilization is notably one of the major challenges on the way to a space-borne gravitational wave observatory. The proposed Laser Interferometer Space Antenna (LISA) is presently under development in an ESA, NASA collaboration. We present a novel method for active laser stabilization and phase noise suppression in such a gravitational wave detector. The proposed approach is a further evolution of the "arm locking" method, which in essence consists of using an interferometer arm as an optical cavity, exploiting the extreme long-run stability of the cavity size in the frequency band of interest. We extend this method by using the natural interferometer arm length differences and existing interferometer signals as additional information sources for the reconstruction and active suppression of the quasi-periodic laser frequency noise, enhancing the resolution power of space-borne gravitational wave detectors.

For the test of general relativity and astrophysical research, measurements of gravitational waves in space are in preparation. Gravitational waves could be detected by observing the strain of space over time. One possible instrument for such measurements is an optical interferometer, which enables comparison of the time-dependent strain in different directions of space. If for such interferometers, the wavelength of the gravitational wave is much bigger than the arm length of the interferometer, then the detected signal can be linearly increased by increasing the interferometer arm length [1,2]. For the LISA mission, the interferometer components are housed in identical satellites $5 \times 10^9$ m apart, forming a triangle in space and providing maximum sensitivity for wave periods between $10^{-4}$ and $10^{-1}$ Hz. The corresponding optical interferometers are of Michelson type with an angle of $\approx 60$ deg between the interferometer arms [1,2]. Figure 1 shows an outline of an example geometry and a nomenclature of the different interferometer arms.

The dimension of such an interferometer implies an increased sensitivity at frequencies below $\approx 30$ mHz [1,2]. The interferometer reference mirrors are quasi-free falling test masses shielded by the spacecraft from non-gravitational disturbances. The spacecraft follow the test mass motion via an electro-mechanical drag-free control system.

Since the motion of the test masses is also subject to interplanetary forces, causing deviations from ideal orbits, the differences in the interferometer arm lengths cannot be kept below a certain limit. In case of LISA, the arm lengths are supposed to breath relatively with an amplitude of about 1% of the mean arm length in the course of a year, causing a slow variation of the corresponding Doppler shifts at a frequency of $\approx 3 \times 10^{-8}$ Hz [1,2] and of the relative angle of $60 \pm 1$ deg.

These unavoidable arm-length differences in space-borne gravitational wave detectors are the source of the detected laser frequency noise in such interferometers at the expected sensitivity maximum at a frequency of $\approx 10^{-3}$ Hz. Due to the unequal arm lengths, the frequency noise is not cancelled as it is the case in ground-based gravitational wave detectors. To overcome this frequency noise, which is several orders of magnitude higher than the desired noise floor [1,2], the post-processing method of time-delay interferometry (TDI) has been developed [3]. Relying upon current space-qualified laser sources, this method leads to high requirements for the time measurement and clock synchronization in the moving spacecraft array as well as information-conserving data reduction (filtering) and interpolation methods or data transfer throughput to earth.

Previously, a method called "arm locking" has been proposed, which in essence is a realization of an optical cavity with the dimension of an interferometer arm, offering frequency noise suppression based on a cavity of the length of one interferometer arm ($L \approx 5 \times 10^9$ m) [4,5]. However, this method does not damp the frequency noise at the frequency corresponding to the inverse round trip time $c/2L$ and multiples of this frequency.

In this paper, we present a method based on this arm-locking approach, which exploits the existing interferometer signals in order to extract information on the time-periodic frequency noise in the arm-locking mode and to remove this noise. The method provides safe and efficient active frequency noise suppression in a naturally self-evident way without influencing the gravitational wave signal, thus causing an enhanced resolution power of space-borne gravitational wave detectors.

In Figure 2, we show the concept of the extended arm-locking method. We restrict the discussion to a two-arm interferometer, but the extension to a three-arm interferometer is straightforward.

A laser on S/C 1 sends light with the phase $P(t)=p(t)+2\pi\nu_0 t$ and the nominal fixed laser frequency $\nu_0$ to S/C 3, where it is responded back to S/C 1. The light received on S/C 1 has the phase $P(t-2L/c)=p(t-2L/c)+2\pi\nu_0(t-2L/c)$. On S/C 1, we can measure the frequency difference

$$(\partial/\partial t)\,[P(t)-P(t-2L/c)]\,/\,(2\pi) = f(t)-f(t-2L/c)$$

between the sent and received light beam. Here, the frequency $f$ is defined by $2\pi f(t) = (\partial/\partial t)p(t)$. Note that $f$ cannot be measured. However, we can measure frequency differences, which are independent of the choice of the nominal $\nu_0$. A heterodyne laser locking is performed between the incoming and sent beam, controlling the sent beam. The filtering function is marked with $G$ in Fig. 2. The indicated function of this feedback loop ideally generates a nearly periodic frequency error $f(t)$, which is repeated for infinite time if the loop works with infinite gain [5]:

$$f(t=t_0) - f(t=t_0 - 2L/c) = 0.$$

The analysis of this opto-mechanical system immediately implies that the periodic frequency error $f(t)$ may be corrected for by adding the negative error to the laser frequency reference signal $f(t-2L/c)$ for the time of $2L/c$, if only the periodic frequency error was known exactly. However, we can send a part of the laser output to S/C 2, receive the reflected beam, and observe the frequency difference

$$\Phi(t_0) = f(t=t_0) - f(t=t_0 - [2L/c - \Delta])$$
$$= f(t=t_0) - f(t=t_0 + \Delta) \approx -\Delta \frac{\partial f}{\partial t}\bigg|_{t=t_0+\Delta/2}.$$

This is an approximation of the derivative of the periodic frequency error function. If we observe the function

$$F(t) = F(t'+\Delta/2) = \begin{cases} -\dfrac{1}{\Delta} \int_0^{t'} \Phi(t'')\,dt'' & 0 < t' < 2L/c \\ 0 & \text{else} \end{cases},$$

we obtain an approximation of the frequency error with respect to the frequency error at time $t = \Delta/2$. If we further compare $F(t) = F(t'+\Delta/2)$ and $f'(t) = f(t'+\Delta/2) - f(\Delta/2)$ and perform a discrete Fourier transformation of $F$ and $f'$ for the period $2L/c$, beginning at time $t = \Delta/2 \neq 0$, it can easily be shown that for ideal arm-locking performance in the frequency domain

$$F(\omega)/f'(\omega) = \text{sinc}(\omega\Delta/2).$$

Thus in principle, we can reconstruct the AC components of the frequency error function $f'(t)$ and subtract it with a time delay of $\xi=2(n+1)L/c$ for the time of $2L/c$, i.e., we subtract the reconstructed $f'(t-\xi)$ from the laser frequency reference signal $f(t-2L/c)$ *in situ*. The value of the integer $n$ may depend on the on-board calculation time needed for reconstruction of $f'$ and should be chosen as low as possible to guarantee little change of the quasi-periodic $f(t)$. After we have subtracted the reconstructed error, we have again a quasi-periodic frequency error, which would be ideally constant and equal $f(\Delta/2)$. Thus under ideal conditions we would completely suppress the laser frequency noise and then begin high-resolution measurements with the self-conserving suppressed noise.

In conclusion, we have presented a method for active suppression of the laser frequency noise, which is desirable for the realization of a space-borne gravitational wave interferometer, enhancing the resolution power of such a detector. The method can be applied under real conditions, where there will be Doppler-shift frequency offsets varying at a frequency of $3\times10^{-8}$ Hz, because the frequency of this variation is out of the measurement band. The Doppler shifts and slow variations can be removed by offset locking and subtraction from the frequency readout $\Phi$. The clock noise of the reference clocks used for heterodyne laser locking can easily be suppressed using the difference frequency between laser lines, locked to the variable-seize cavity. The arm-length information $\{L, c\Delta\}$, which is needed for the correction procedure, can be obtained by observing correlations of time-shifted high-frequency components of $\Phi$. We have seen that in first order perfect cancellation of the laser frequency noise can be reached. The efficiency of the method will be limited by time measurement errors, which will be very small with respect to a typical round-trip time difference of $|\Delta|\approx0.3$ s, i.e., a negligible time shift error will be reached with standard clocks. Moreover, small errors in the determination of $\Delta$ become negligible if the procedure described in this paper is repeated from time to time, since the small relative error is multiplied with itself and reduced each time the procedure is performed. For increased performance, we may repeat the subtraction of weighted fractions of the corrections by multiplying the correction functions with an optimized $\alpha(\omega)$. Simulation calculations for the arm-locking mode were already performed by other groups [4,6]. We performed simulation calculations for the extended method presented in this paper with standard noise input up to the heterodyne bandwidth in the kilohertz range [6], which prove that stable operation and a permanent laser noise suppression of several orders of magnitude is reached simultaneously at relevant frequencies below, at and above $c/2L$.

**Acknowledgments**

The author wishes to thank Dr. Hans-Reiner Schulte and Dr. Ulrich Johann for helpful discussions.

**References**

[1] K. Danzmann et al., "LISA-proposal for a laser-interferometric gravitational wave detector in space", Max-Planck-Institut für Quantenoptik, Report MPQ 177 (1993).

[2] The European Space Agency, "LISA Laser Interferometer Space Antenna, a cornerstone mission for the observation of gravitational waves", System and Technology Study Report, ESA-SCI(2000)11, July 2000.

[3] M. Tinto and S. Dhurandhar, "Time-Delay Interferometry", arXiv.org:gr-qc/0409034 (2004).

[4] B.S. Sheard, M.B. Gray, D.E. McClelland, and D.A. Shaddock, "Laser frequency stabilization by locking to a LISA arm", *Phys. Lett. A* **320**, 9 (2003).

[5] M. Tinto and M. Rakhmanov, "On the laser frequency stabilization by locking to a LISA arm", arXiv:gr-qc/0408076 v1 (2004).

[6] J. Sylvestre, "Simulations of laser locking to a LISA arm", arXiv:gr-qc/0408055 (2004).

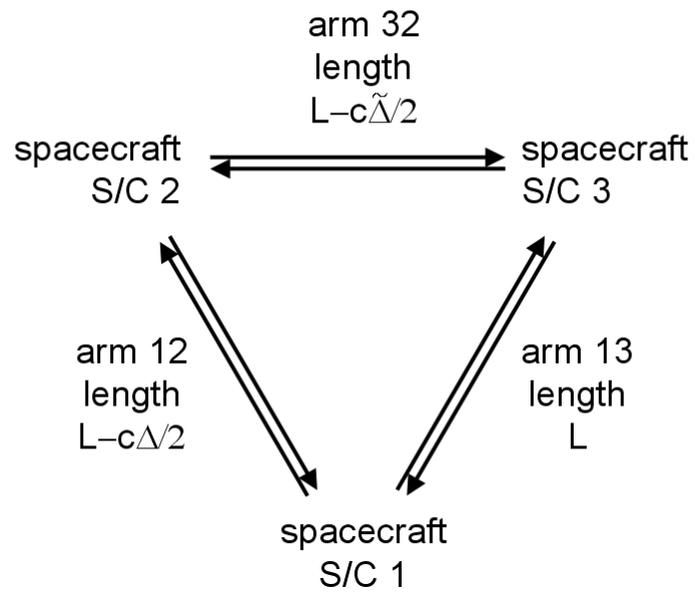

**Fig 1.** Outline of an example geometry with a nomenclature of the three identical spacecraft and interferometer arms (see also Refs. [1,2]).

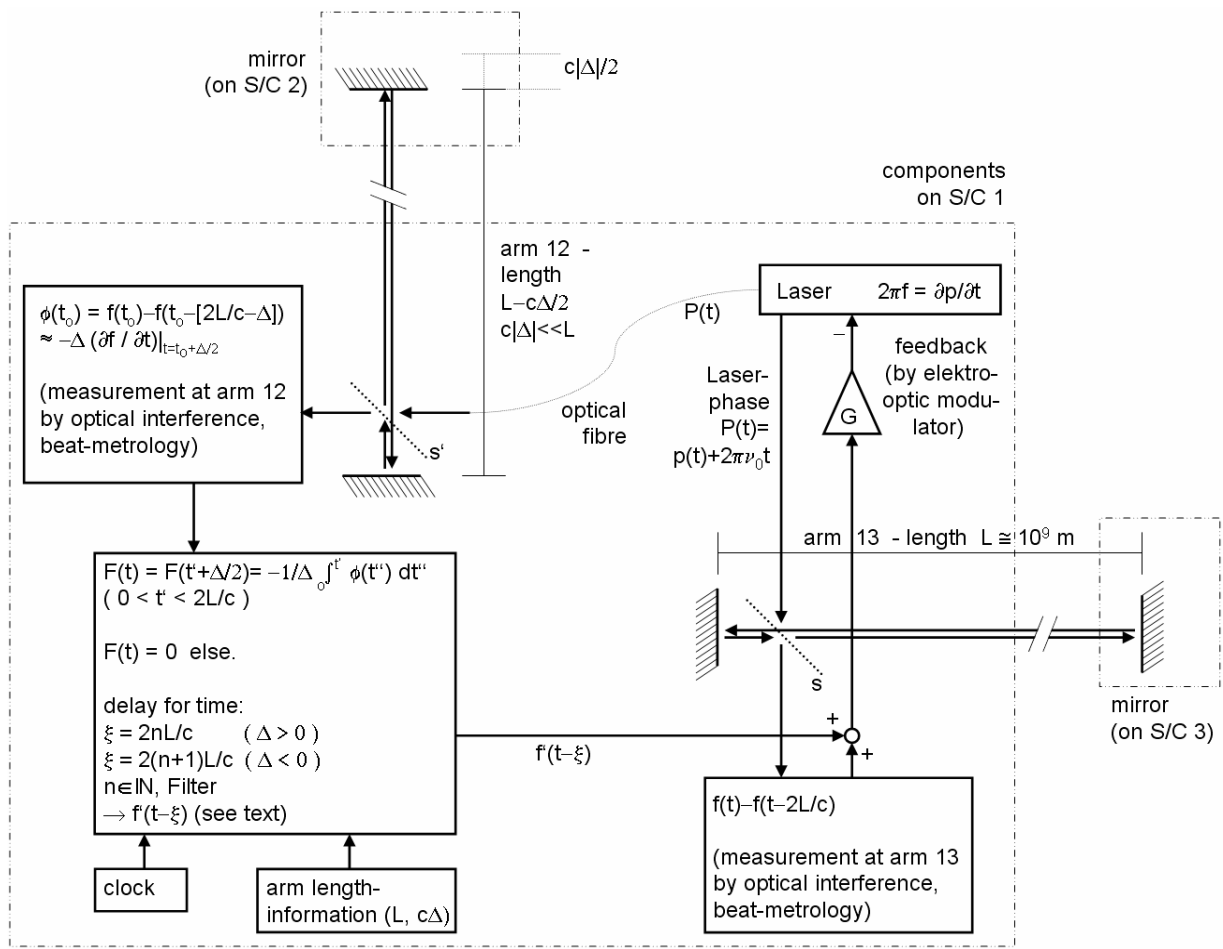

**Fig 2.** Principle of the laser stabilization method. A heterodyne arm-locking configuration is maintained between S/C 1 and the mirror on S/C 3. The laser beam is sent to S/C 2 in parallel and the frequency difference $\Phi$ measured on S/C 1 is used to reconstruct and subtract the quasi-periodic frequency error $f(t)$. Here s and s' symbolize optical units containing beam-splitters. See also Fig. 1 for the nomenclature.